\documentclass[letterpaper,12pt]{article}   
\usepackage{osajnl2} 
  \usepackage{amssymb}  
\def\AOM {acousto-optic modulator}
\def\CCD {charge-coupled device}

\def\fl {fluorescence}
\def\MOT {magneto-optical trap}

\def\wrt {with respect to}

\def\ARC {Australian Research Council} 

\newcommand{\degrees}{$^{\circ}$}
\newcommand{\degC}{$^{\circ}$C}

\newcommand{\si}{$\sim$}
\newcommand{\microns}{$\mu$m}
\newcommand{\uK}{$\mu$K}

\newcommand{\ee}[1]{\ensuremath{\times 10^{ #1}}}

\newcommand{\fastT}{$^{1}S_{0}-\,^{1}P_{1}$}    
\newcommand{\coolingT}{$^{1}S_{0}-\,^{3}P_{1}$}  

\newcommand{\Yb}{$^{171}$Yb}	
\newcommand{\Ybtwo}{$^{172}$Yb}
\newcommand{\Ybthree}{$^{173}$Yb}
\newcommand{\Ybfour}{$^{174}$Yb}
\newcommand{\sigsig}{$\sigma^{+}-\sigma^{-}$}

\begin{document}


\title{Sub-Doppler cooling of ytterbium with the \fastT\ transition including \Yb\ (I=$\frac{1}{2}$)}


\author{N. Kostylev,  E. Ivanov, M. E. Tobar, and J.J. McFerran$^{*}$}
\address{$^1$School of Physics, University of Western Australia, 6009 Crawley, Australia}
\address{$^*$Corresponding author: john.mcferran@uwa.edu.au}

\begin{abstract}  
We report on the  
sub-Doppler laser cooling of neutral \Yb\ and \Ybthree\ in a magneto-optical trap using the \fastT\ transition at 398.9\,nm.  We use two independent means to estimate the temperature of the atomic cloud for several of the Yb isotopes. 
The two methods of MOT-cloud-imaging and release-and-recapture show consistency with one another.
  Temperatures below 400\,\uK\ and 200\,\uK\ are recorded for \Yb\ and \Ybthree, respectively, while \si1\,mK is measured for both \Ybtwo\ and \Ybfour.  By comparison, the associated  1D Doppler cooling temperature limit is 694\,\uK.  
  The sub-Doppler cooling of the $I=\frac{1}{2}$ \Yb\ isotope in a \sigsig\ light-field trap adds further evidence that the Sisyphus cooling mechanism is occurring in such 3D \MOT s. 

\end{abstract}


\maketitle 


\section{Introduction}

Particular attention is being given to the cooling and trapping of ytterbium at present since it has proven to be  
 useful for a range of atomic studies and applications.   Ytterbium, as the quantum absorber for optical lattice clocks, has shown remarkable results\cite{Hin2013, Lem2009a} and is the atom of choice for a number of atomic clock projects\cite{Koh2009a,Par2013}.  Yb was  used to make the first spin-singlet Bose-Einstein condensate~\cite{Tak2003b}, a consequence of it being a two-electron atom.  Simultaneous multi-isotope trapping has been performed~\cite{Lof2001}, culminating in photo-association  
spectroscopy of excited hetero-nuclear ytterbium molecules~\cite{Bor2011}.  
Multi-species magneto-optical trapping with Yb has been demonstrated~\cite{Oka2010, Mun2011, Bor2013}, and recently, quantum degenerate mixtures involving Yb have been produced~\cite{Har2011a,Han2011}.  
Furthermore, cold Yb also appears to be a favourable candidate for exploring short-range interactions\cite{Wol2007}, studying artificial Abelian and non-Abelian gauge potentials~\cite{Ger2010a}, and searching for a permanent  electric-dipole moment~\cite{Rat2013}.  This short list demonstrates the popularity and versatility of  the laser cooled ytterbium atoms. 

Group II  atoms have a reputation for having higher than expected temperatures in magneto-optical traps\cite{Xu2002, Lof2000,Mar2003b}: only infrequently is the Doppler cooling temperature limit reached in single-stage cooling\cite{Xu2003a,McF2010} (where single-stage refers to the use of a single transition).  Although single-stage cooling with the heavy group II atoms using the \fastT\ transition does not produce exceptionally low temperatures,  the temperature reached at the early stage of cooling is important, since lower temperatures lead to higher transfer rates to proceeding stages 
(for example,  into a \coolingT\  \MOT\ (MOT) or a far off-resonance optical trap).   
  Here we present, what we believe to be, the lowest temperatures produced when magneto-optically trapping ytterbium with the $(6s^{2})\,^{1}S_{0}-(6s6p)\,^{1}P_{1}$    line at 398.9\,nm. 

A number of previous schemes have been employed in the trapping of neutral Yb. Some examples are:   (1) direct loading of a \fastT\ \MOT\  from a effusive source \cite{Lof2001}, (2) loading into a \coolingT\ MOT with a Zeeman slower using the \fastT\ line~\cite{Kuw1999}, and (3) loading with a Zeeman slower and two stage cooling in the MOT~\cite{Koh2009a,Par2013}.  Here we report on results obtained with a Zeeman slower and  \fastT\ line cooling in a \MOT.  
 We show data supporting the sub-Doppler cooling of the \Yb\ and \Ybthree\ isotopes using  $^{1}S_{0} (F=\frac{1}{2})-\,^{1}P_{1}  (F'=\frac{3}{2})$ and  $^{1}S_{0} (F=\frac{5}{2})-\,^{1}P_{1}  (F'=\frac{7}{2})$ transitions, respectively.   The sub-Doppler cooling of \Yb\ is notable, since Yb has a nuclear spin of $\frac{1}{2}$ and thus the total angular momentum quantum number $F=\frac{1}{2}$ for the ground state.  Therefore, the motion-induced orientational sub-Doppler cooling process~\cite{Dal1989a,Chu1989a}, which requires $F_{g}\geq1$, cannot be present in the \sigsig\ light field configuration used here.  We provide further evidence that the sub-Doppler temperatures observed arise from the Sisyphus cooling mechanism usually ascribed to the lin $\perp$ lin (or $\pi^{x} \pi^{y}$) light field configuration~\cite{Mar2003b}.  This is deduced from the knowledge that a 3-dimensional arrangement of  \sigsig\  beams can produce Sisyphus type polarisation gradients~\cite{Hop1997,Sch1999a}. 

\section{Magneto-optical trap and Zeeman slower}

The vacuum system for the cooling and trapping of Yb comprises an effusion cell, a Zeeman slower section, the main trapping chamber and vacuum pumps.  The pressure in the main chamber is maintained at $\lesssim$4\ee{-7}\,Pa with the use of a getter pump and a dual getter plus ion pump (Saes getter).  The former is located near the entrance to the Zeeman slower and the latter, 30\,cm beyond the centre of the main chamber (in the direction of the atom flow). 
 The Yb atom source is a custom made effusion cell (Createc, GmbH) designed from suitable components that  do not bind or alloy with Yb; 
  for example, most of the crucible is made from tantalum, while the nozzle is composed of 1\,mm diameter, 20\,mm long stainless steel tubes.  
 The oven has the capability of heating the Yb sample to well over 1000\,\degC; however, here we typically operate at 400-410\,\degC.  
   The oven consumes 10.6\,W at 400\,\degC.   The effusion cell's controller temperature concurs well with direct measurements made on the atomic beam (further discussion about the temperature appears below).   
  
 Light from a Ti:sapphire laser is frequency doubled in a resonant cavity occupying a 12\,mm long, Brewster-cut, lithium triborate crystal ($\theta=90$\degrees, $\phi=31.7$\degrees).  The crystal is maintained at 42\degC. Approximately 140\,mW of 398.9\,nm light is produced by the cavity, although, for most measurements here the cavity was sideband-locked with a power output of \si90\,mW.

\begin{figure}[h]
 \begin{center}
{		
  \includegraphics[height=8cm,keepaspectratio=true]{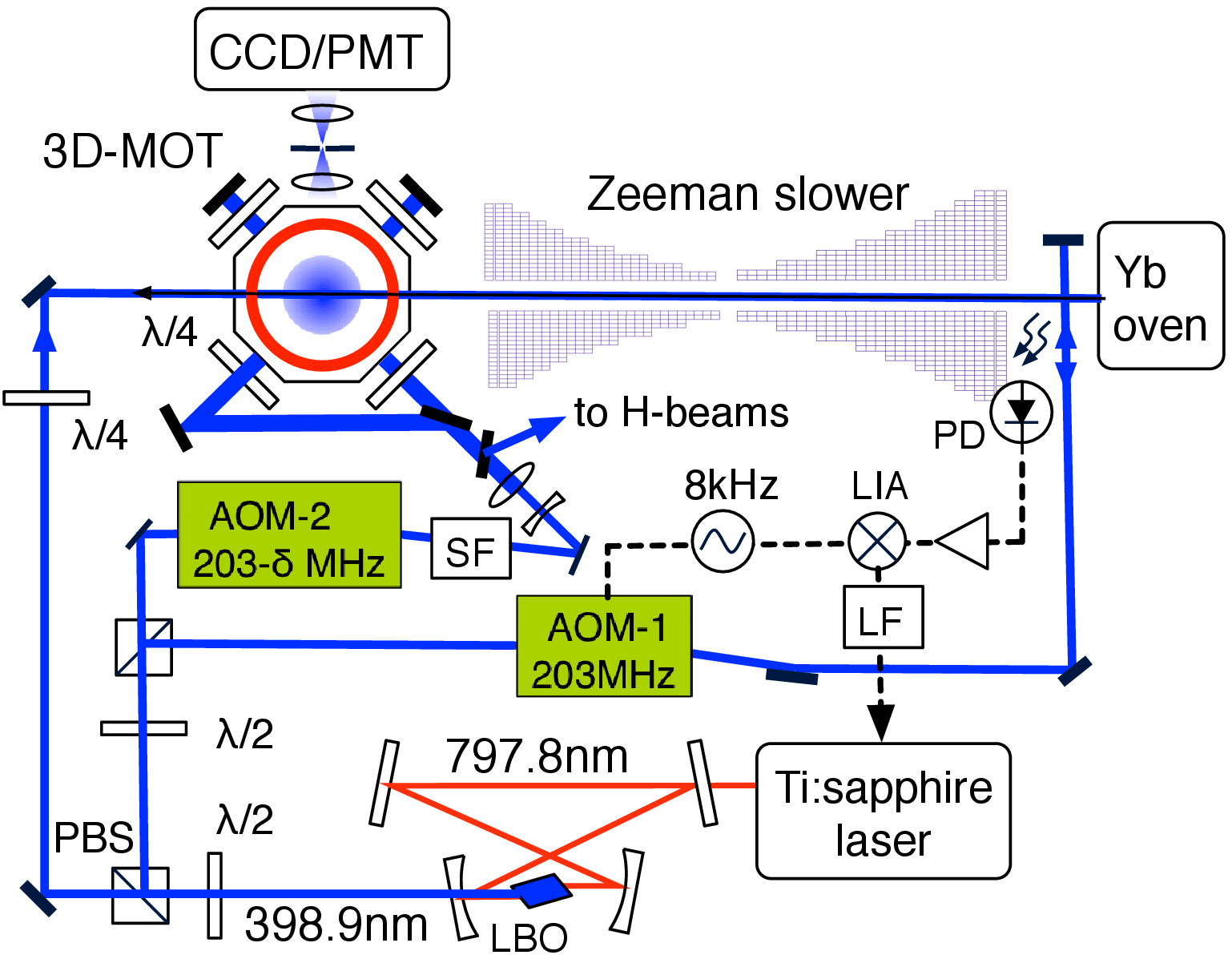}} 
\caption[]{\footnotesize      (Color online) Main components of the experimental set up. The light from a frequency doubled Ti:sapphire laser is divided into three beams: one beam  is used in the Zeeman slower, another is the laser light for the six beams of the \MOT\ (MOT) and a third is used to lock the laser to the  \fastT\  Yb transition.  
Frequency modulation spectroscopy is used for the laser stabilisation via probing of the atomic beam 9\,cm from the exit of the Yb oven. A 8\,kHz sinusoid drives  an acousto-optic modulator  (AOM-1) and the corresponding modulation is imprinted on the fluorescence collected by a photodiode (PD) adjacent to a viewport.  The demodulation of the PD signal via a lock-in amplifier (LIA) generates a correction signal that is suitably filtered and sent to the Ti:sapphire laser. AOMs provide the required frequency detunings for the Zeeman slower and the MOT beams.  Definitions:  $\delta$, frequency detuning;  H-beams, the horizontal beam of the MOT and its retro-reflection,  LBO, lithium triborate crystal; LF, loop filter (including integral gain  with a 10\,Hz servo bandwidth); $\lambda/2$ ($\lambda/4$), half (quarter) wave plate;   PBS, polarising beam splitter; PMT, photo-multiplier tube (cell); and SF, spatial filter.  
  } \label{ExpSetUpYb}
\end{center}
\end{figure}

We use a typical magneto-optical trap (MOT) configuration with two anti-Helmholtz coils producing the quadrupole magnetic field, along with three pairs of orthogonal retro-reflected beams in the $\sigma^{+}-\sigma^{-}$ arrangement \cite{Raa1987,Ste1991}.  The strong axis of the MOT lies in the horizontal plane.   It is a compact design with 52\,mm separating the centres of the current carrying coils. 
 The coil separation distance to coil diameter ratio is 0.85 and 58 windings are used in each coil.  With 6.2\,A passing through the windings a field gradient of 0.31\,T\,m$^{-1}$ (or 31\,G\,cm$^{-1}$) is estimated to be produced along the strong axis.  The heat dissipated is removed using water cooling through   copper tubing in close thermal contact with the coils. 
 The vacuum chamber walls immediately surrounding the atoms in the MOT are of titanium. It is hard enough to form reliable con-flat flanges and it is free from any ferrous materials that may otherwise generate unwanted magnetic field gradients.   The titanium component is a single piece with an octagonal  profile in the vertical plane.  Each face has provision for  con-flat flange mounting (the bore diameter for each MOT arm is 16\,mm).  Windows bonded to stainless steel (316)  flanges attached to the Ti chamber allow passage for the MOT beams: two in  the vertical plane, and one horizontal.   Optics, such as quarter wave plates and mirrors, are attached directly to the Ti chamber. 
 We use free space coupling of light through the MOT chamber;  
 though  the design of the 
  Ti chamber  ensures a high degree of orthogonality between the MOT arms.   

The magneto-optical trap is fed with a 28\,cm long Zeeman slower based on a two-stage tapered coil with a zero crossing magnetic field~\cite{Cou2003} and a circularly polarised  laser beam tuned -7\,$\Gamma$ from the \fastT\ transition centre ($\Gamma$ is the natural linewidth of $2\pi\times28.9\times10^{6}$\,rad\,s$^{-1}$).  
  The inner most coil diameter is 22\,mm.  The maximum field strength of the Zeeman slower has not been measured inside the central tube, but a calculation based on the full winding configuration  suggests that the mean $B$-field gradient is approximately 0.18\,T\,m$^{-1}$ (18\,G\,cm$^{-1}$), with the $B$-field reaching 0.025\,T (250\,G) at the ends of the Zeeman slower when 10\,A flows through the coils (we have the provision for using higher current through the two outer sets of windings if later needed).   
The exit of the Zeeman slower is 12.5\,cm from the centre of the MOT. 
For optimum MOT loading we find the optical power ratio between the MOT and Zeeman slower to be $3:2$, where the power at the MOT is for the six beams. 



Viewports shortly after the exit of the effusion cell provides access for  laser locking to the \fastT\ transition of the atom beam. Line broadening from the 
velocity spread is minimised  by having the probe beams at right angles to the atomic beam (the linewidth is approximately 77\,MHz indicating an atomic beam divergence of \si60\,mrad~\cite{Dem2008}). We use frequency modulation spectroscopy to lock to the centre of the transition. The locking beam is derived from the main output of the frequency doubling cavity, but is first positively frequency shifted by 203\,MHz  ($7\,\Gamma$) with the use of an \AOM\ (AOM-1 in Fig.~\ref{ExpSetUpYb}).  The modulation for the FM spectroscopy makes use of this AOM. 
For frequency control in the MOT, a separate beam  is passed through a second AOM  (AOM-2) and  the difference frequency between the two AOM RF signals provides the frequency detuning, $\delta$, for the MOT.  
The AOM-2  creates a defect in the beam profile (in our case) and the beam from the SHG cavity is slightly irregular.  To remedy this we use a spatial filter that produces a spherical-Gaussian beam in the fundamental mode. The spatial filter comprises a 75mm focal length lens and a 25\,\microns\ diameter pin hole; 75-80\,\% of the light is throughput.  We found the spatial filter to be
necessary for producing sub-Doppler temperatures with \Ybthree, though not for \Yb. 

With regard to the Yb oven, we probed the escaping beam of atoms with a 398.9\,nm beam aligned at 45\degrees\ (at the location of the MOT chamber) with and without the Zeeman slower acting. The shift in the peak of the line profiles (most clearly seen with \Ybtwo\ and \Ybfour) was 560\,MHz. From this we deduce that the maximum of the modified Maxwell-Boltzmann velocity distribution occurs at \si320\,m\,s$^{-1}$, and the corresponding temperature is 420\degC~\cite{Bud2008}.  Similarly we found that the most probably velocity of the slowed atoms, with the Zeeman slower operating, was less than 15\,m\,s$^{-1}$.   With regard to the MOT, the ratio of relative  strengths of the fluorescence signals from the trapped atoms is $ 0.26:0.72:0.10:1$ for the isotopes: 171 through to 174, respectively.  In comparison, the ratio of   the relative strengths of the relevant transitions is:   $ 0.36:0.69:0.28:1$~\cite{Dei1993}. 


\begin{figure}[h]
 \begin{center}
{		
  \includegraphics[height=6cm,keepaspectratio=true]{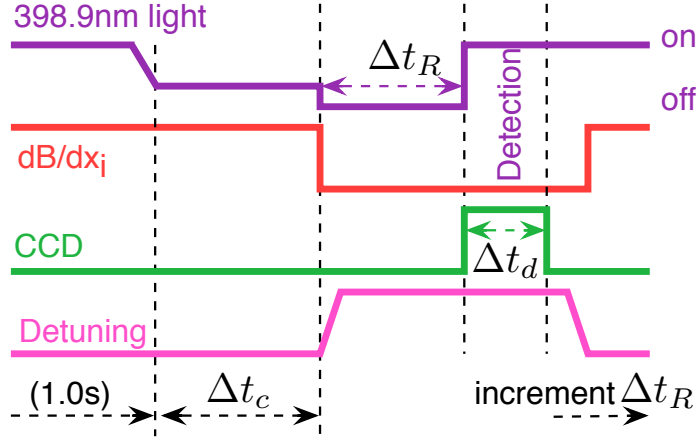}} 
\caption[]{\footnotesize      (Color online)  The sequence for magneto-optical trapping and detection for temperature measurements of Yb.  The 398.9\,nm light is used for atom cooling and slowing. The intensity is reduced to \si10\,\% of the loading intensity over  $\Delta t_{c}$  prior to the atoms' release ($\Delta t_{c}=10$\,ms unless otherwise indicated).    The $B$-field gradient in the MOT is simply switched off and on.   The exposure time, $\Delta t_{d}$, for the CCD is 1\,ms and the  laser detuning for detection is normally set to $-0.8\,\Gamma$.  $\Delta t_{R}$ is the duration of free expansion.  
  } \label{SequenceYb}  
\end{center}
\end{figure}

\section{Temperature and  cloud size}

While there are many aspects of a \MOT\ that could be tested, here we focus on temperature, 
since this appears to be where the most significant difference occurs compared to previous work.
The temperature of the cold atom cloud in the \MOT\ is determined by two means: the first using CCD imaging of the MOT cloud and the second using  the release and recapture method\cite{Chu1985,Let1988}.
 In the first  case, the ballistic expansion of the atomic cloud is recorded in a series of images with a \CCD\ (CCD), where the object-plane is perpendicular to gravity.  For each subsequent image, the MOT is reloaded and the free expansion time of the atoms is extended.  The control of the fields and frequency detuning is represented in Fig.~\ref{SequenceYb}, which also indicates three time intervals:  $\Delta t_{R}$, representing the free expansion period of the atoms; $\Delta t_{d}$, the exposure time of the CCD camera; and $\Delta t_{c}$, a period of reduced light intensity to further cool the atoms.  Here we use $\Delta t_{d}=1$\,ms. The free expansion duration commenced at 0.5\,ms and was incremented in 0.5\,ms steps for successive images.  We assume that molasses restricts further cloud expansion during the exposure period (if there is some further residual expansion then our measurements provide an upper-estimate of the cloud temperature).   
A root-mean-square (rms) radius of the cloud is deduced from each image using two orthogonal profiles, and 
from the sequence of cloud radii,   curve fitting by use of  $r^{2}(t)= r^{2}_{0}+(k_{B}T/M)t^{2}$  evaluates the temperature, $T$, and the initial (or steady-state) rms cloud size, $r_{0}$; 
 $k_{B}$ is the Boltzmann constant and $M$ is the atomic mass. 
 The atom cloud is imaged onto the CCD with two pairs of convex lenses, each pair having identical focal lengths, producing a magnification of one.   
 An iris is placed between the pairs of lenses to filter out undesired scattered light.   The pixel size of the CCD is 6.5\,$\mu$m and is not deemed to affect the cloud size measurements (given that the rms cloud radii are more  than 30 times greater than this).

\begin{figure}[h]
 \begin{center}
{		
  \includegraphics[height=8cm,keepaspectratio=true]{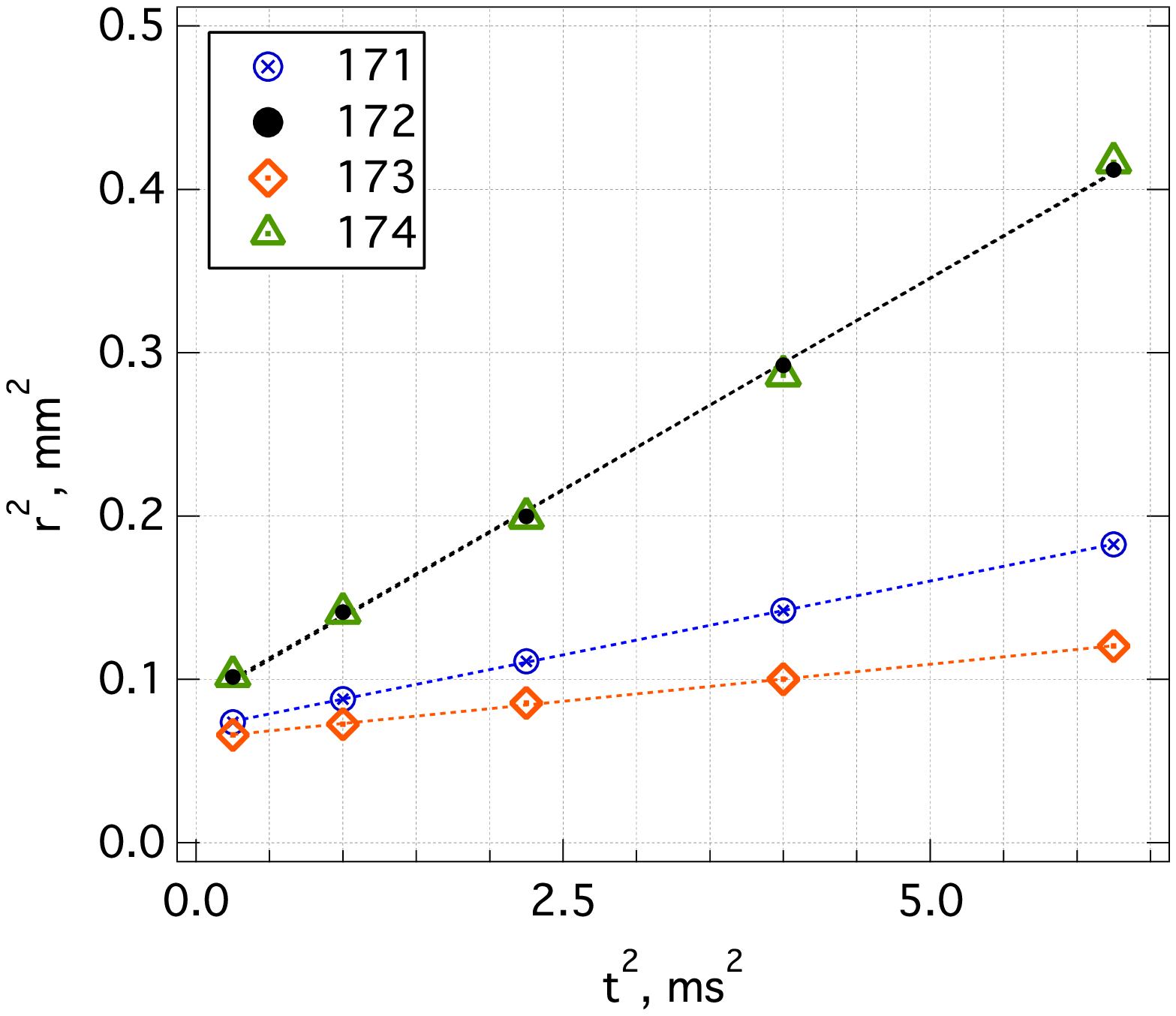}} 
\caption[]{\footnotesize      (Color online)     Atom cloud radius (rms) as a function of the release time from the MOT for four of the Yb isotopes.  From the slopes the temperatures are: 380\,\uK\ (\Yb), 1.08\,mK (\Ybtwo),   190\,\uK\ (\Ybthree) and 1.09\,mK\ (\Ybfour).  

  } \label{BallisticTraces}
\end{center}
\end{figure}

    Examples of the atom cloud size (rms radius) versus the free expansion time are displayed in Fig.~\ref{BallisticTraces} for isotopes 171 through to 174. Here the corresponding temperatures are: 380\,\uK\ (\Yb), 1.08\,mK (\Ybtwo),   190\,\uK\ (\Ybthree) and 1.09\,mK\ (\Ybfour). Based on a series of repeated measurements, the statistical 1-$\sigma$  uncertainty is   \si8\,\% for both the fermions  and bosons.  The difference in temperatures between the bosonic isotopes (\Ybtwo\ and \Ybfour)   and the fermionic isotopes (\Yb\ and \Ybthree) is manifest.  These data were recorded with a MOT  beam radius $(e^{-2})$  of $w\approx$6\,mm, apart from \Ybthree, where $w=$3.2\,mm.  
   In the case of \Yb, where $r_{\mathrm{max}}\approx0.4$\,mm from Fig.~\ref{BallisticTraces}, the  
   underestimation of the cloud size is \si1\,\% based on the selection (or finite-size) effect of the 2D-Gaussian beam profile of the detection beam. 
   This could produce temperature estimates \si3\,\% lower than what they otherwise should be. We have accounted for this effect in the measurements below (being more important when the MOT beams, and therefore detection beams, are reduced in diameter)

\begin{figure}[h]
 \begin{center}
{		
  \includegraphics[height=8cm,keepaspectratio=true]{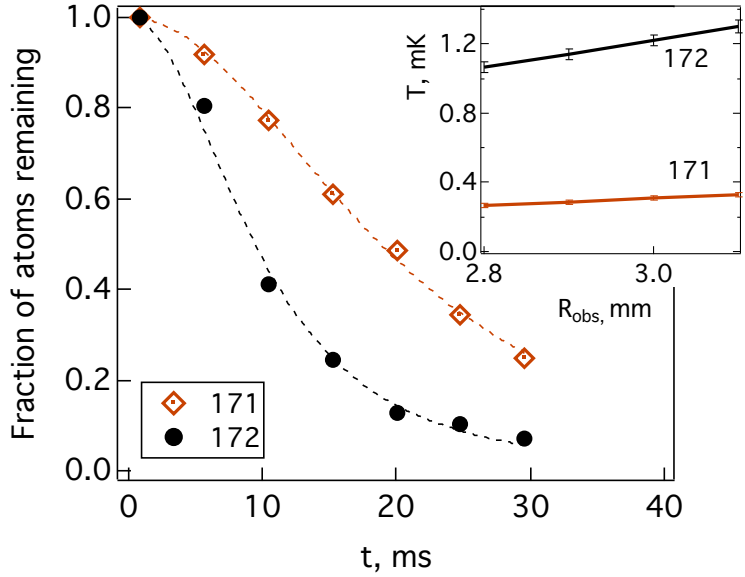}} 
\caption[]{\footnotesize      (Color online)  
 Atom fraction remaining versus free expansion time for isotopes  $^{171}$Yb and \Ybtwo.  
 The dotted line is a modelled fit with the observation radius $R_{\mathrm{obs}} (\mathrm{rms}) =3.0$\,mm producing $T=310$\,\uK\ and 1.2\,mK for \Yb\ and \Ybtwo\, respectively. 
 The inset shows the influence of $R_{\mathrm{obs}}$ on the derived temperature. 

  } \label{RandR}
\end{center}
\end{figure}

 For the release and recapture method, in place of the CCD a photo-multiplier cell (Hamamatsu 10492-001) was used to detect the atoms'  fluorescence,  and the aperture that is used to reduce stray light was expanded so as to have minimal affect on the observation volume.  Again a sequence was employed in which the free expansion time was extended for each fluorescence recording.  Here the release time was incremented in steps of 4\,ms.   
Fig.~\ref{RandR}  shows the fraction of atoms remaining in the observation zone set by the intersection of the MOT beams, versus the release time from the MOT. 
 The results for two isotopes, \Yb\ and  \Ybtwo\   are shown; the corresponding temperatures are $310^{+30}_{-60}$\,\uK\ and $1.2^{+0.2}_{-0.3}$\,mK, respectively, assuming an observation volume with an effective rms radius of  $3.0^{+0.2}_{-0.4}$\,mm.  
   An uncertainty arises with this technique in relation to the observation radius, $R_{\mathrm{obs}}$, which defines the volume from which \fl\ is collected and imaged onto the photo-multiplier cell. The spatial filtering of the MOT beam  produces a beam profile with well defined dimensions, but the observation volume extends beyond the intersection of the three beams (and into the arms) and so needs to be taken into account. 
 We still treat this extended volume as having an effective radius, despite the observation volume not being spherical, to enable a curve fit to the data.  
 Although we have omitted the detail here the effective rms radius of 3.0\,mm is consistent with a total observable volume of the imaging system being slightly smaller than that defined by the inner walls of the chamber. 
 The inset Fig.~\ref{RandR}  shows the dependence of temperature on $R_{\mathrm{obs}}$ for \Yb\ and  \Ybtwo.  
  The modelled fit does not take into account gravity, which is a fair approximation to make given the temperatures observed here.
 Despite the uncertainty in temperature that $R_{\mathrm{obs}}$ generates,    the difference in temperature between the fermionic \Yb\ and bosonic \Ybtwo\ is again very evident. 
 


\begin{figure}[h]
 \begin{center}
{		
  \includegraphics[height=10cm,keepaspectratio=true]{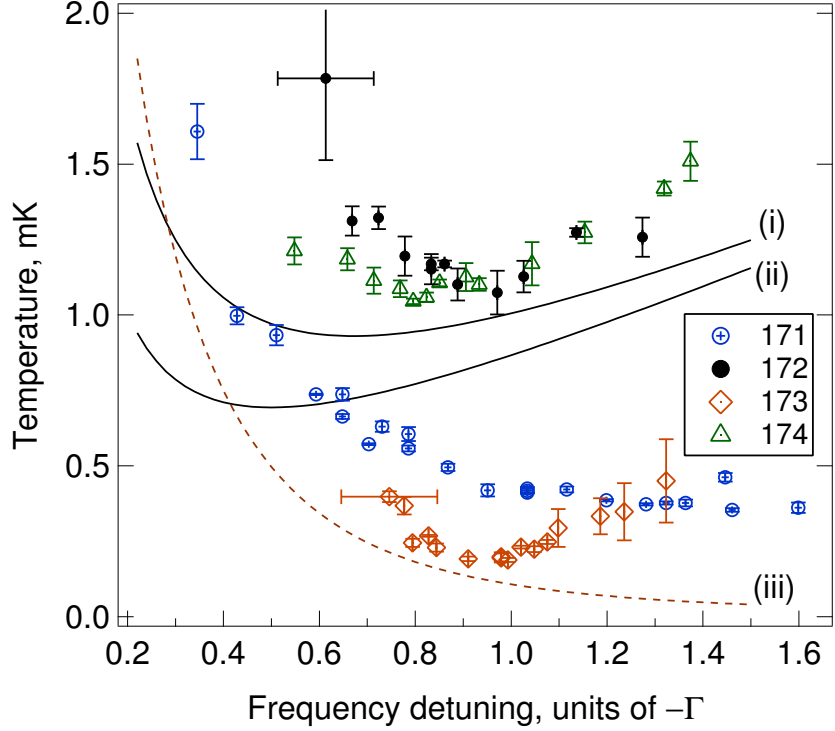}} 
\caption[]{\footnotesize      (Color online)  Atom cloud temperature versus  frequency detuning (red) for isotopes:  $^{171}$Yb, $^{172}$Yb, $^{173}$Yb  and $^{174}$Yb. $\Gamma = 2 \pi \times 28.9$\,MHz.  The solid lines show the cooling limit from Doppler cooling theory for intensities of (i)  $s_{0}=0.8$ and (ii) $s_{0}=0$.  The dashed line (iii) is a prediction from 1D-sub Doppler cooling theory based on the motion-induced orientational cooling (corkscrew) mechanism for $F_{g}=1\leftrightarrow F_{e}=2$ and $s_{0}=0.2$. 
  } \label{TemprvsdetuningG46}
\end{center}
\end{figure}

    Measurements of temperature versus frequency detuning  have been carried out for the isotopes 171 through to 174 using the cloud imaging method, with
 the results shown in Fig.~\ref{TemprvsdetuningG46}.  The uncertainties shown by the error bars are statistical, produced by the  line fitting used for the $r^{2}$ versus $t^{2}$  measurements. 
 Both the B-field gradients and the 398.9\,nm light are switched off for the free expansion.  
 The intensity is reduced to \si0.2$s_{0}$ before the release of the atoms ($s_{0}=I/I_{S}$ is the on-resonance saturation parameter, where $I_{S}=595$\,W\,m$^{-2}$ is the saturation intensity for the \fastT\ line).  
 The solid lines in Fig.~\ref{TemprvsdetuningG46} are the theoretical cooling limits from standard Doppler cooling theory\cite{Let1989} for the cases where the intensity $s_{0} = 0$ (trace ii) and  $s_{0} = 0.8$ (trace i), the latter being included to show the affect of intensity.  
  While the temperatures of the bosonic isotopes remain above the Doppler cooling limit, $T_{D}$, of 694\,$\mu$K, those of  $^{171}$Yb and \Ybthree\ appear well below this limit (for $|\delta|>0.6\,\Gamma$ in the case of \Yb), demonstrating the presence of  a sub-Doppler cooling mechanism (the frequency detuning range  of Fig.~\ref{TemprvsdetuningG46} is somewhat restricted because AOM-2 is limited to single pass operation).
  
   While the relative frequency detuning for the individual isotopes has a low measurement uncertainty, there is a greater level of uncertainty between isotopes, since  the lock point on the various discriminating slopes may be slightly different.  The uncertainty on the absolute frequency detuning is estimated to be $0.2\,\Gamma$. 
  We point out that the temperature of the \Ybthree\ appears particularly sensitive to the shape of the beam profile of the cooling light. Prior to use of the spatial filter, the \Yb\  cloud reached temperatures similar to those obtained without the filter; 
  however, those of \Ybthree\ were above 700\,\uK.  
  We also note that the temperatures recorded so far remain well above the recoil limiting  temperature of \si700\,nK for the \fastT\ transition.

Previous reports on 398.9\,nm cooling of Yb indicated temperatures of 0.7\,mK for several of the Yb isotopes~\cite{Par2003} (with no difference in temperature between the fermionic \Yb\ isotope and other bosonic isotopes).  There is also a report of \si2\,mK for $^{174}$Yb~\cite{Lof2000}, demonstrating that achieving Doppler limiting temperatures with Yb on the violet line is not straightforward.    
 In relation  to the temperature of Yb,  the focus is generally on the 
 second stage of cooling using the \coolingT\ transition, so the reports on the \fastT\ line cooling of Yb appear limited. Some groups avoid the 398.9\,nm MOT cooling altogether and load with the \coolingT\  transition, using  the \fastT\ for Zeeman slowing only~\cite{Hon2002}.   Results here hold promise for improved loading efficiency into subsequent cooling and trapping stages.   

From the analysis of Ref.~\cite{Dal1989} in relation to the \sigsig\ configuration, we can perform an approximate calculation of the equilibrium temperature of the atomic cloud for sub-Doppler cooling.  
  We may compare this to the case of \Ybthree,  \fastT\ ($F_{g}=\frac{5}{2}-F_{e} = \frac{7}{2}$), despite the calculation being for an $F_{g}=1$ to $F_{e} =2$ transition~\cite{comment1}.  Using an intensity of 0.2\,$I_{S}$, with a corresponding Rabi frequency of 1.1\ee{8}\,rad\,s$^{-1}$, the estimated temperature versus detuning is shown by the dashed line in Fig~\ref{TemprvsdetuningG46}.   Although we do not expect to get precise agreement here, we see that model and experiment do not differ too greatly   (the effect of AOM-2's beam deflection is see on the \Ybthree\ temperature as $\delta$ increases).  

Sub-Doppler cooling is not permitted via the motion-induced orientational cooling mechanism in atoms with $F_{g}=\frac{1}{2}$ when using the $\sigma^{+}-\sigma^{-}$  configuration (irrespective of dimensionality)~\cite{Dal1989}.   
However, investigations of the 
 electric field polarisation structure of three-dimensional light fields reveal that  with the use of the \sigsig\ configuration, Sisyphus polarisation gradients can occur~\cite{Mol1991,Hop1997}, and hence the Sisyphus cooling mechanism may be expected to occur.  This appears to be the most likely explanation for the sub-Doppler cooling exhibited here for \Yb~\cite{Mar2003b}. 
 The lower temperature of \Ybthree\  \wrt\ \Yb\ is not too surprising, since the former has a nuclear spin of $\frac{5}{2}$ and the relevant transition is  $^{1}S_{0} (F=\frac{5}{2})-\,^{1}P_{1}  (F=\frac{7}{2})$; hence,  the expected  motion-induced orientational  cooling should occur (where an imbalance in magnetic sub-state populations brought about by the atom's motion and optical pumping causes different absorption rates between beams of opposing direction and helicity).  Given that the Sisyphus mechanism occurs for \Yb, it is   
 anticipated to occur for \Ybthree\ as well; hence, the two cooling mechanisms occur simultaneously to produce greater damping and lower temperatures.   Note, a difference in cloud temperatures between fermionic isotopes of the same element has been observed before in a \sigsig\ MOT~\cite{Wal1994,McF2010}.    
 Assuming Sisyphus cooling is present, then we can make comparisons with predictions by M\o lmer  where the temperature versus detuning was evaluated for four different polarisation configurations in 3D~\cite{Mol1991}.   Based on calculations for the lin$\perp$lin configuration and $F_{g}=1$ to $F_{e} =2$, the predicted temperature falls from \si50\,\uK\ to \si20\,\uK\ for the detuning range considered here.  We may treat this as a lower bound on the temperatures that could be produced via experiment on the \fastT\ transition.  
 

%

\begin{figure}[h]
 \begin{center}
{		
  \includegraphics[height=8cm,keepaspectratio=true]{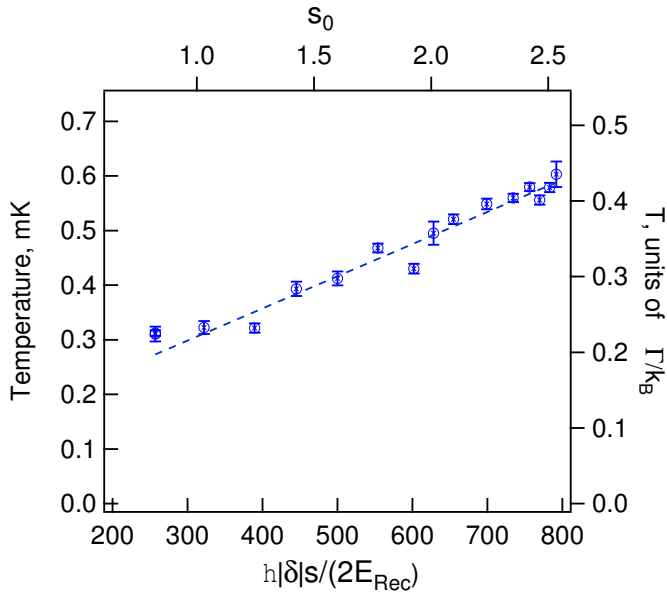}}    
\caption[]{\footnotesize      (Color online) Temperature versus intensity for \Yb\ at a fixed frequency detuning of -1.4$\Gamma$.  The saturation parameter $s= s_{0}/(1+(2\delta/\Gamma)^{2})$ 
  } \label{Tvsintensity}  
\end{center}
\end{figure}

\begin{figure}[h]
 \begin{center}
{		
  \includegraphics[height=8cm,keepaspectratio=true]{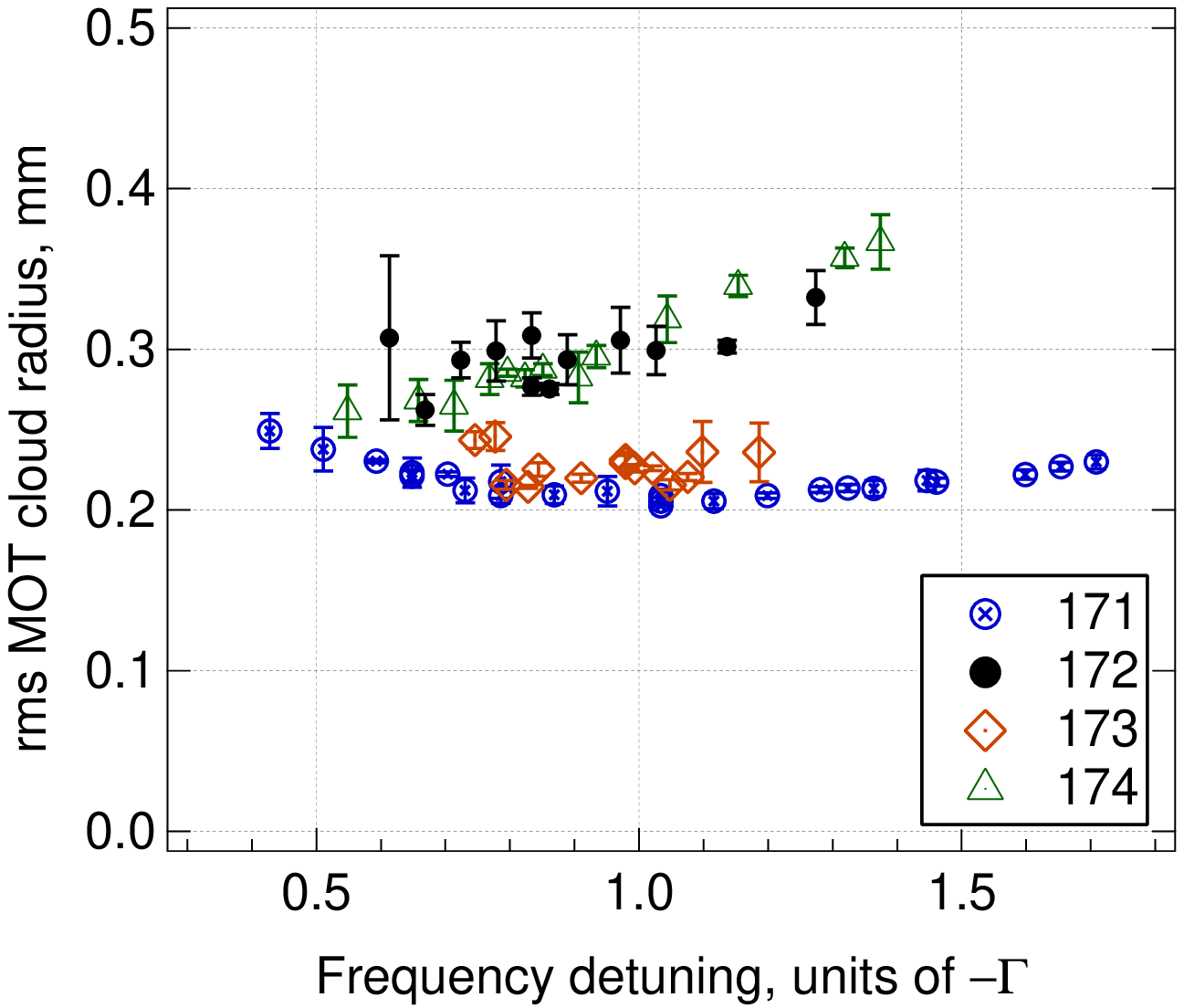}}    
\caption[]{\footnotesize      (Color online) Atom cloud rms radius ($e^{-1/2}$) versus  frequency detuning, $|\delta|$, for isotopes:  $^{171}$Yb through to $^{174}$Yb. $\Gamma = 2 \pi \times 28.9$\,MHz. 
  } \label{radius}
\end{center}
\end{figure}

We have also carried out temperature versus intensity measurements, as shown in Fig.~\ref{Tvsintensity}. 
 For these measurements the intensity is changed in  the $\Delta t_{c}=15$\,ms period prior to the release of the atoms. The same intensity is used during the MOT loading to maintain roughly equal numbers of atoms for each measurement.
Figure~\ref{Tvsintensity} shows the \Yb\ cloud temperature versus intensity for a fixed detuning of $\delta = -1.4\Gamma$.  The upper abscissa shows the intensity normalised by the saturation intensity of 595\,W\,m$^{-2}$, while the lower abscissa is the intensity in terms of the quantity  $\hbar |\delta| s/2E_{\mathrm{Rec}}$ to allow comparison with theoretical estimates~\cite{Cas1995a}.  $E_{\mathrm{Rec}} = (\hbar k)^{2}/2M$ is the recoil energy, and $k$ is the wavenumber. 
 The saturation parameter $s= s_{0}/(1+(2\delta/\Gamma)^{2})$.  
The temperatures here are considerably lower than those predicted by 1D Doppler cooling theory, which for $\delta = -1.4\Gamma$ predicts a minimum temperature of 1.5\,mK at $s_{0}=0.5$.  
Following Castin and M\o lmer~\cite{Cas1995a} (who consider the lin $\perp$ lin configuration) a universal relationship between temperature and intensity in a MOT can be described where $3k_{B}T/2 = a (\hbar |\delta|s/2) +bE_{\mathrm{Rec}}$.  The parameter $a$ depends on the total angular momentum, $J$ (or $F$), and is typically smaller for larger values of $J_{g}$.  From this work we find the coefficient $a=2.5\pm0.4$  for \Yb, where the uncertainty takes into account those in $\delta$ and $s$.  
Our value for $a$ is noticeably smaller than that obtained by Maruyama \emph{et al.}~\cite{Mar2003b}, but here we satisfy $s<1$ ($s_{\mathrm{max}}=0.28$).  
 In the work of Ref.~\cite{Mar2003b} the properties of Yb in a MOT were investigated, highlighting  the temperatures produced using the \coolingT\ transition at 556\,nm ($T_{D}=4.4$\,\uK).   This showed clear a difference in  
 $T$ versus $s_{0}$ behaviour between the bosonic and fermionic isotopes suggesting the presence of a sub-Doppler cooling mechanism, but unlike here, did not manifestly show temperatures below $T_{D}$ ($T_{D} = 694$\,\uK\ for \fastT).  

From the free expansion measurements of the MOT cloud  we also infer the  equilibrium state cloud size, i.e., the same measurements used to find the temperature are also used to find the initial rms cloud radius ($e^{-1/2}$).  
 The rms  radius is plotted as a function of frequency detuning in Fig.~\ref{radius}.  Within the range of $\delta$ tested there is not a great deal of variation in cloud size for a given isotope.  Between the isotopes we see that the bosonic isotopes have a slightly larger cloud size.   From the equipartition theorem the spring constant of the trap $\kappa\propto T/r^{2}$.
 For the bosons and \Yb, $\kappa$ is similar in magnitude, while it is slightly lower for \Ybthree.
 So far, we do not see the effects of radiation trapping, where re-absorbed fluorescence limits the density of the atoms in the MOT.   We estimate the maximum density of the atoms to be just over $10^{9}$\,cm$^{-3}$.   Increasing the density of the atoms in the MOT will be the focus of future work.

\section{Conclusions}

We have shown sub-Doppler cooling of ytterbium isotopes \Yb\ and \Ybthree\ (fermions) and efficient cooling of \Ybtwo\ and \Ybfour\ (bosonic) by use of the \fastT\ transition in a \MOT.  Measurements have been conducted using two methods, albeit, one with lower uncertainty.  The temperature for the fermionic isotopes appear to be the lowest yet reported  for this single stage of cooling using the \fastT\ line, with temperatures well below 400\,\uK\ for \Yb\ and $T\lesssim 200$\,\uK\ for \Ybthree.  We attribute the low temperatures to the design of the vacuum chamber that ensures a high degree of othogonality between the MOT beams, as well as the titanium housing surrounding the atom cloud, which minimises irregularities in the quadrupole magnetic field.   
There is also a distance of 80\,cm between the effusive source and the location of the MOT, which may lessen the disruption from remaining high speed atoms emitted by the source on the cold atom cloud. 

Since \Yb\ has a nuclear spin of $\frac{1}{2}$, the ground state has $F=\frac{1}{2}$; therefore, according to 1-D sub-Doppler cooling theory there is no mechanism to generate atom  cloud temperatures below the Doppler limiting temperature, $T_{D}$, for the \sigsig\ configuration used here.
  We have demonstrated that sub-Doppler cooling is present for the \Yb\ isotope, which supports the analyses in Refs.~\cite{Hop1997,Sch1999a} that polarization gradients of the Sisyphus type are present in 3D systems.  We note that indications of sub-Doppler cooling were observed previously for the spin $\frac{1}{2}$ isotope of $^{199}$Hg~\cite{McF2010} and with the \coolingT\ line in Yb~\cite{Mar2003b}.  Here we show for the first time atom cloud temperatures well below the Doppler limiting temperature, $T_{D}$, for a nuclear spin $\frac{1}{2}$ atom.
   Hence, we show further evidence that the Sisyphus mechanism is acting in \sigsig\ \MOT\ arrangements. 

\section{Acknowledgements}

This work was supported by the \ARC\ (LE110100054).  J.M. is supported through an ARC Future Fellowship (FT110100392).  We thank Gary Light and the UWA Physics workshop staff for their technical expertise.  We are grateful to J\'{e}r\^{o}me Lodewyck and Andrew Ludlow  for their advice on experimental design aspects.  We thank S\'{e}bastien Bize and Rodolphe Le Targat for helpful discussions. 


\end{document}